\documentclass[aps,prl,twocolumn,groupedaddress,showpacs,floatfix]{revtex4}
\usepackage{graphicx}
\usepackage{psfig}
\usepackage{epsf}
\usepackage{epsfig}

\begin{document}

\title{Relaxation to non-equilibrium in expanding ultracold neutral plasmas}

\author{T.\ Pohl}
\author{T.\ Pattard}
\author{J.M.\ Rost}

\affiliation{Max Planck Institute for the Physics of Complex Systems,
N{\"o}thnitzer Str.\ 38, D-01187 Dresden, Germany}

\date{\today}

\begin{abstract}
We investigate the strongly correlated ion dynamics and the degree of coupling
achievable in the evolution of freely expanding ultracold neutral plasmas.
We demonstrate that the ionic Coulomb coupling parameter $\Gamma_{\rm i}$
increases considerably in later stages of the expansion, reaching the strongly
coupled regime despite the well-known initial drop of $\Gamma_{\rm i}$ to
order unity due to disorder-induced
heating. Furthermore, we formulate a suitable measure of correlation and show that $\Gamma_{\rm i}$ calculated from the ionic temperature
and density reflects the degree of order in the system if it is
sufficiently close to a quasisteady state.
At later times, however, the expansion of the plasma cloud becomes faster
than the relaxation of correlations, and the system does not reach
thermodynamic equilibrium anymore.
\end{abstract}


\maketitle

Freely expanding ultracold neutral plasmas (UNPs) \cite{Kil99}
have attracted wide attention both experimentally
\cite{Kul00,Kil01,Rob00,Rob04} and theoretically
\cite{Rob02,Kuz02,Maz02,PPR04,Tka01}.
A main motivation of the early experiments was the
creation of a strongly coupled plasma, with the Coulomb coupling parameter
(CCP) $\Gamma = e^2/(a k_{\rm B} T) \gg 1$ (where $T$ is temperature and $a$ is
the Wigner-Seitz radius). From the experimental setup of
\cite{Kil99}, the CCPs of electrons and ions were estimated to be of the
orders of $\Gamma_{\rm e} \approx 30$ and $\Gamma_{\rm i} \approx 30000$,
respectively. By changing the frequency of the ionizing laser, the electronic
temperature can be varied, offering the prospect of controlling the coupling
strength of the electrons and creating UNPs where either one, namely the
ionic, or both components could be strongly coupled.

However, due to unavoidable heating effects \cite{Rob02,Bon97,Mur01} these
hopes have not materialized yet, and only $\Gamma_{\rm e} \approx 0.2$ and
$\Gamma_{\rm i} \approx 2$ have been confirmed.  Furthermore, the evolution of
the expanding plasma turns out to be a rather intricate
problem of non-equilibrium plasma physics for which a clear definition of the
degree of correlation is not obvious to begin with.

The goal of this letter is twofold: Firstly, we will formulate a consistent
measure of correlation for expanding ultracold plasmas, and secondly we
demonstrate that the strongly correlated regime with $\Gamma_{\rm i} \approx 10$
for the ionic plasma component can be reached by simply waiting until the plasma
has (adiabatically) expanded long enough under already realized experimental
conditions. This is remarkable in the light of alternatives proposed to
increase $\Gamma_{\rm i}$
\cite{Mur01,Ger03,Kil03,PPR04a,PPR04b,Kuz03}
which are experimentally rather involved.

Substantiating both of our statements theoretically requires the ability
to propagate the plasma numerically over a long time with full account of the
ionic correlations. To this end, we have developed a hybrid molecular dynamics
(H-MD) method \cite{PPR04} for the description of ultracold neutral plasmas.
In our approach, ions and recombined atoms are propagated in the
electronic mean-field potential with the full ion-ion interaction taken
into account. The much faster and weakly coupled electrons, on the other hand,
are treated on a hydrodynamical level. Elastic as well as inelastic
collisions, such as three-body recombination and electron-impact ionization, are
incorporated using a Monte-Carlo procedure \cite{Rob03,PPR04b}.
The H-MD approach accurately describes the strongly coupled ionic
dynamics and therefore allows us to realistically study the plasma relaxation
behavior for long times.

\begin{figure}[bt]
\centerline{\psfig{figure=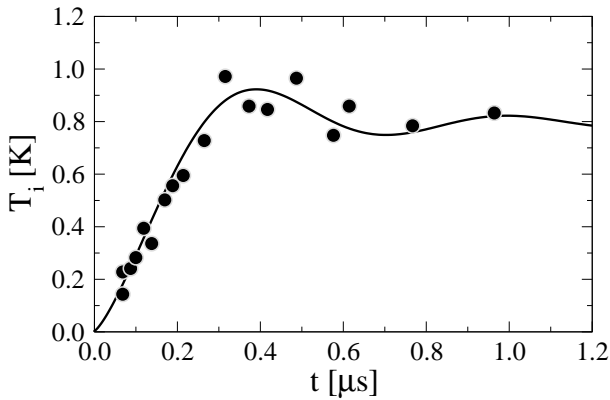,width=8cm}}
\caption{\label{fig1}
Calculated ion temperature (solid line) for a plasma of $10^6$ Sr ions with
initial peak density $\rho_0(0) = 2 \cdot 10^9$cm$^{-3}$ and electron
temperature $T_{\rm{e}}(0)=38$K, compared to experimental results (dots)
\cite{Chen04}.
The fact that the
experimental ion number is about a factor of ten larger than in our
calculation does not affect the time evolution of the ionic temperature, since
there is no significant adiabatic ion cooling on the timescale considered in
figure \ref{fig1}.}
\end{figure}

Assigning $\Gamma_{\rm i}$ for an expanding plasma by extracting a temperature
from the kinetic energy of all ions is complicated by the fact that the radial
expansion contributes considerably to this energy \cite{Kil04}. In our approach,
we can determine a {\em local} temperature from the ion velocity components
perpendicular to the (radial) plasma expansion \cite{PPR04}.
Additionally, the distribution of thermal velocities of {\em all} plasma ions
is found to be well described by a Maxwell-Boltzmann distribution
corresponding to an average temperature $T_{\rm i}$ even at relatively early
times. Experimentally, the time evolution of the average ion temperature is
determined from the corresponding Doppler broadening of optical
transition linewidths \cite{Kil04,Chen04}. The close agreement between
experiment \cite{Chen04} and theory (figure \ref{fig1}) supports both the
experimental scheme of
extracting an ionic temperature as well as the assignment of a temperature to
the transversal ion velocities in the H-MD approach.

Remarkably, the initial relaxation of the average ion temperature exhibits
temporal oscillations, in contrast to the known behavior of weakly coupled
plasmas. For the latter, the timescale $t_{\rm corr}$ of the initial
build-up of ion-ion correlations is typically much smaller than the timescale
$t_{\rm rel}$ for the relaxation of the one-particle distribution function.
Based on this
so-called Bogoliubov functional hypothesis, which is one of the fundamental
concepts in kinetic theory \cite{Kli82}, the different
relaxation processes can be separated, resulting in a monotonic behavior of
the correlation energy (and hence the ion temperature) \cite{Mor98}. Molecular
dynamics simulations of the relaxation behavior of homogeneous one-component
plasmas show that the ion temperature starts to undergo damped oscillations
around its equilibrium
value if both of these timescales become equal, which happens for
$\Gamma_{\rm{i}}(0) \agt 0.5$ \cite{Zwi99}. Therefore, the nonmonotonic ion
relaxation observed in ultracold plasmas may be seen as a direct manifestation
of the violation of Bogoliubov's hypothesis.

Compared to the homogeneous plasmas considered in \cite{Zwi99}, the oscillations
of the average ionic temperature damp out much quicker in the present case.
This can be attributed to the fact that the Gaussian density
profile of the UNPs created in current experiments leads to a spatial
dependence of the correlation timescale $t_{\rm corr}$, the build-up of
correlations being fastest in the center of the plasma where the density is
highest, and becoming slower towards the edge of the plasma cloud. As a
consequence, the local ionic temperature shows not only
temporal, but also pronounced spatial oscillations, which however tend to
become averaged out if the spatial average over the whole plasma cloud is taken.

\begin{figure}[tb]
\centerline{\psfig{figure=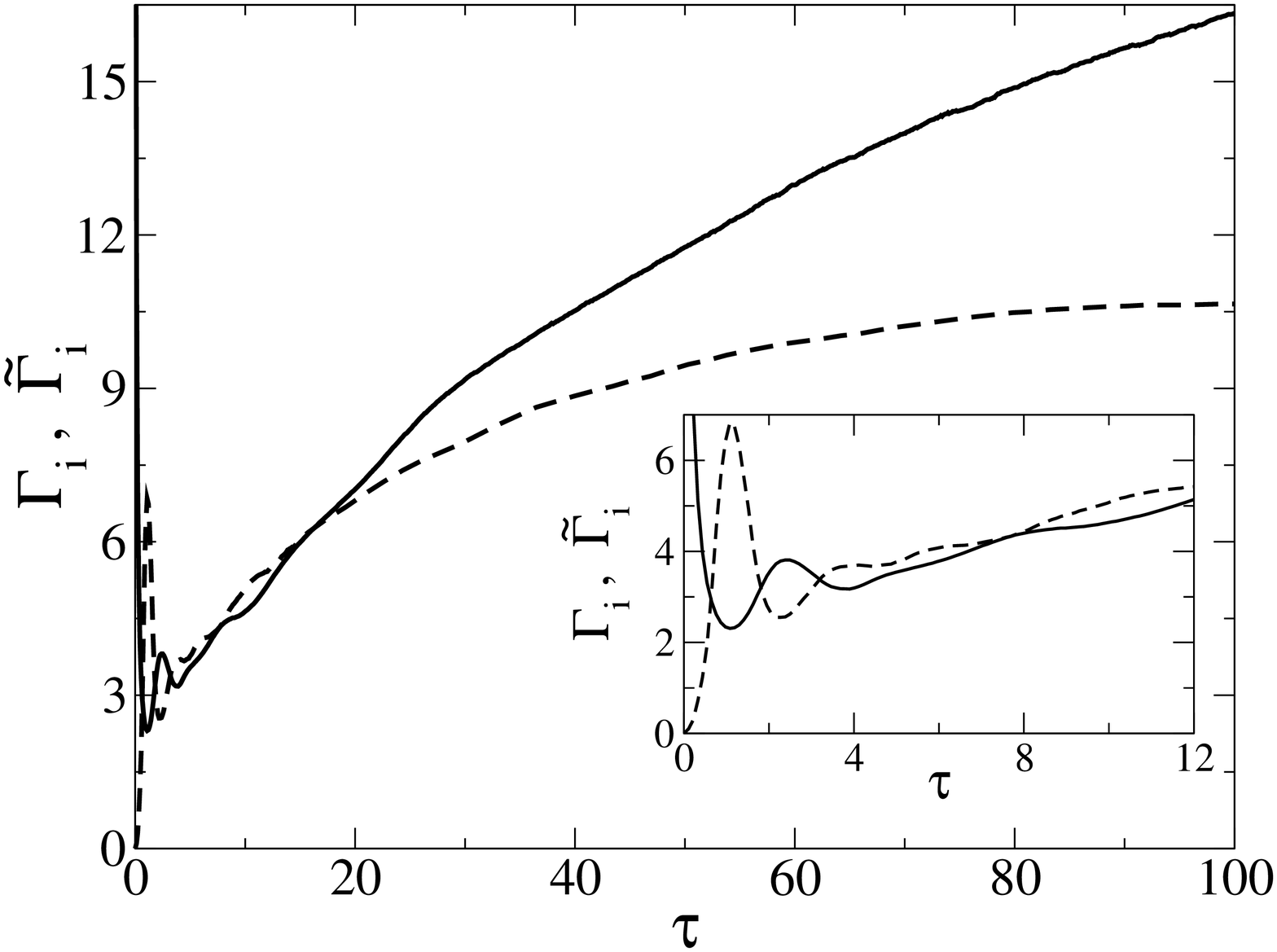,width=7.8cm}}
\caption{\label{fig2}
Ionic Coulomb coupling parameter for a plasma with $N_{\rm{i}}(0)=5\cdot10^4$,
$\bar{\rho}_{\rm{i}}(0)=1.1\cdot10^9$cm$^{-3}$ and $T_{\rm{e}}(0)=50$K.
The solid line shows the CCP calculated from the average temperature and
density, the dashed line marks the CCP extracted from
pair correlation functions (see text). Inset: blow-up of the short-time
behavior.}
\end{figure}
Having established the approximate validity of assigning a global temperature
to the plasma ions, it becomes possible to define a corresponding CCP
$\Gamma_{\rm i}$. While the initial ion relaxation reveals some interesting
strong-coupling effects as discussed above, disorder-induced heating
\cite{Mur01,Kuz02} drives the ion component to the
border of the strongly coupled fluid regime $\Gamma_{\rm{i}}\approx2$ and
therefore limits the amount of correlations achievable in UNPs. However, so far
this could be verified only for the early stage of the plasma evolution
\cite{Mur01,Kuz02,Kil04}. The present H-MD approach allows us to study also
the long-time behavior of the ion coupling.

In figure \ref{fig2}, we show $\Gamma_{\rm i}$ (solid line) as a
function of $\tau=\omega_{\rm p,0}t$ for a plasma with $N_{\rm{i}}(0)=5\cdot10^4$,
$\bar{\rho}_{\rm{i}}(0)=1.1\cdot10^9$cm$^{-3}$ and $T_{\rm{e}}(0)=50$K,
determined in a central sphere with a
radius of twice the root-mean-square radius of the expanding
plasma. (In the following, dimensionless units are used where time is scaled with the initial  
plasma frequency $\omega_{\rm{p,0}} = \omega_{\rm{p}}(t=0)$ and $\omega_{\rm{p}}
= \sqrt{4 \pi e^2 \bar{\rho}_{\rm i} / m_{\rm i}}$.)
As can be seen in the inset, $\Gamma_{\rm i}$ quickly drops down to
$\Gamma_{\rm{i}}\approx2$. After this initial stage, however, 
$\Gamma_{\rm i}$ starts to increase again due to the adiabatic cooling of
the ions during the expansion.
Indeed, CCPs of more than 10 are
realized at later stages of the system evolution, showing that cold
plasmas well within the strongly coupled regime are produced with the present
type of experiments.

Neglecting the influence of the changing
correlation energy as well as inelastic processes, the adiabatic law for the
plasma expansion \cite{Dor98} yields $T_{\rm i}
\bar{\rho}_{\rm{i}}^{-2/3} = \mbox{const.}$. Hence, $\Gamma_{\rm i}$ should
increase $\propto \bar{\rho}_{\rm i}^{-1/3}$ as the plasma expands, ultimately leading to
coupling strengths of $10^2$ or even larger at very long times.
For a classical plasma in
thermodynamical equilibrium, the Coulomb coupling parameter is a direct
measure of the amount of correlations, and properties such as
pair correlation functions etc.\ can be parametrized by this single quantity.
However, the UNPs created in the present type of experiments are
non-equilibrium systems. Initially, e.g., they are created in a completely uncorrelated state,
so that the high value of $\Gamma_{\rm i}$ caused by the ultralow
temperature of the ions has no relation at all with the correlation properties
of the system. At later times, the system relaxes towards a local 
equilibrium. However, the plasma is
freely expanding, and hence constantly changing its steady state. Thus, the
plasma is in a non-equilibrium state at all times, and one must ask to what
extent $\Gamma_{\rm i}$ really parametrizes the correlations present in the
plasma.

\begin{figure}[tb]
\centerline{\psfig{figure=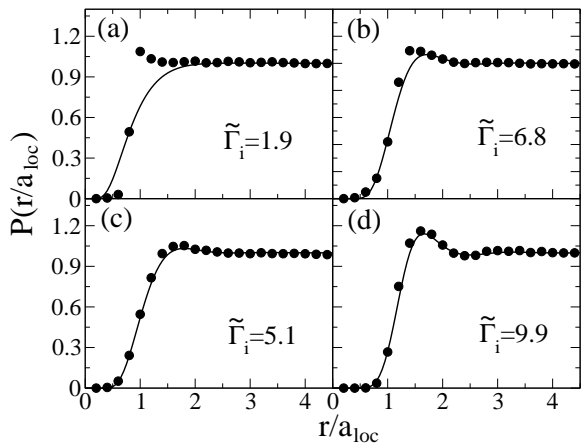,width=7.6cm}}
\caption{\label{fig3}
``Pair correlation functions'' of the plasma of fig.\ \ref{fig2} at four
different times $\tau= 0.54$, $\tau=1.1$, $\tau=10$ and $\tau=60.7$. The
$\tilde{\Gamma}_{\rm i}$
indicated in the figure is obtained by fitting the distribution of scaled
interionic distances (dots) with pair correlation functions for a homogeneous
plasma given in \cite{Hel86} (solid line).
}
\end{figure}

To this end, we compare $\Gamma_{\rm i}$ as obtained
above with an alternative value $\tilde{\Gamma_{\rm i}}$ (dashed line in
figure \ref{fig2}) parametrizing correlation properties of the plasma. As in
\cite{PPR04a}, we have calculated the distribution $P(r/a_{\rm loc})$ of 
interionic distances rescaled by the local Wigner radius. These distribution functions
are fitted to the known pair correlation function
$g(r/a,\tilde{\Gamma}_{\rm{i}})$ of an equilibrium plasma given in \cite{Hel86}
(figure \ref{fig3}). From the fit,
a value $\tilde{\Gamma_{\rm i}}$ is extracted at several times.
As can be seen in figure \ref{fig3}, at very early times the distribution of
scaled interionic distances is not very well fitted to a pair correlation
function of a homogeneous plasma in equilibrium. Again, this is due to the
fact that the system is far away from its steady state, and a
single parameter does not describe the correlation properties of the plasma
in an adequate way. However, the interionic distances quickly relax, and they
are well described by a pair correlation function of an equilibrium system at
later times.
Hence, we conclude that the value of $\tilde{\Gamma_{\rm i}}$ is suitable
for parametrizing the correlation properties of the plasma cloud once it came
sufficiently close to equilibrium, and that it indeed reflects the
degree of coupling in the plasma.

Comparing $\Gamma_{\rm i}$ and $\tilde{\Gamma_{\rm i}}$ in figure \ref{fig2},
several conclusions can be drawn. As discussed above, and has been well known
before, in the very early phase of the system evolution there is no relation
between $\Gamma_{\rm i}$ and $\tilde{\Gamma_{\rm i}}$ since the plasma is too
far away from equilibrium.  As the plasma relaxes towards this equilibrium,
$\Gamma_{\rm i}$ and $\tilde{\Gamma_{\rm i}}$ rapidly
approach each other, showing that during this stage $\Gamma_{\rm i}$ is a good measure for the correlation properties of the ions. In particular,
the correlations building up in the system are indeed those of a strongly
coupled plasma with a CCP well above unity. Moreover, the transient
oscillations characteristic of the relaxation process which are apparent in
$\Gamma_{\rm i}$ also appear in $\tilde{\Gamma_{\rm i}}$, however with a
``phase shift'' of $\pi$. This phase shift is due to the fact that a minimum
in the temperature means a maximum in $\Gamma_{\rm i}$ for a given density.
Since total energy is conserved, a minimum in the thermal kinetic energy
corresponds to a maximum in the potential energy, i.e.\ to an increased number
of pairs of closely neighboring ions, and therefore to a pair correlation
function with enhanced probability for small
distances and consequently a minimum in $\tilde{\Gamma_{\rm i}}$.

At later times, both curves diverge again and the plasma evolves back towards an
undercorrelated state. At first sight, this seems very surprising since the
plasma should relax towards equilibrium rather than away from it. However,
as argued above, the plasma is freely expanding and the corresponding
equilibrium properties are constantly changing. We interpret figure \ref{fig2}
as being again evidence for the break-down of the Bogoliubov assumption of a
separation of timescales, in this case
of the correlation time $\tau_{\rm corr}$ and the hydrodynamical
timescale $\tau_{\rm{hyd}}$, i.e.\ the characteristic time for the plasma
expansion.

The timescale $\tau_{\rm hyd}$ may be determined from the
relative change of macroscopic plasma parameters, such as the ion temperature
or density. Due to the transient oscillations of the ion temperature we
choose the ion density to characterize the change of the plasma
properties (other choices such as, e.g., $a \propto \bar{\rho}_{\rm i}^{-1/3}$ lead to the
same conclusions since they result in a simple constant proportionality factor 
$1/\alpha$ of order unity in the expression for $\tau_{\rm{hyd}}$). Then
\begin{equation} \label{thyd}
\tau_{\rm{hyd}} \approx \frac{1}{\alpha} \frac{{\bar{\rho}_{\rm
i}}}{\dot{\bar{\rho}}_{\rm i}}=
\frac{1}{\alpha} \left(1+
\frac{\tau^2}{\tau_{\rm{exp}}^2}\right)\frac{\tau_{\rm{exp}}^2}{3\tau} \; ,
\end{equation}
where we have used the selfsimilar solution for the collisionless quasineutral
plasma expansion \cite{Dor98} with $\tau_{\rm{exp}}=\sigma(0)\omega_{\rm p,0}
\sqrt{m_{\rm{i}}/(k_{\rm{B}}T_{\rm e})}$.
On the other hand, binary correlations are known to relax on the timescale of
the inverse of the plasma frequency in the strongly coupled
regime \cite{Zwi99} for an initially uncorrelated state, and somewhat slower
if the initial state already exhibits spatial ion correlations
\cite{Mur01}, $\tau_{\rm{corr}} \agt \omega_{\rm{p,0}}/\omega_{\rm{p}}$.
The selfsimilar plasma expansion then yields
\begin{equation}
\tau_{\rm{corr}}=\left(1+\tau^2/\tau_{\rm{exp}}^2\right)^{3/4} \; .
\end{equation}
Therefore, $\tau_{\rm corr}$ is initially much smaller than $\tau_{\rm hyd}$,
but ultimately exceeds $\tau_{\rm hyd}$ as the plasma expands,
leading to an inevitable break-down of the Bogoliubov condition.
Consequently, the build-up of correlations in the system cannot follow the
changing equilibrium anymore, and correlations freeze out as indicated by the
leveling-off of $\tilde{\Gamma_{\rm i}}$ towards a constant value.

\begin{figure}[tb]
\centerline{\psfig{figure=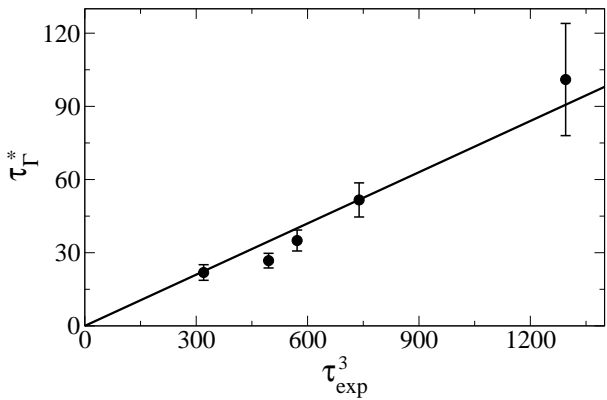,width=8cm}}
\caption{\label{fig4}
$\tau_{\Gamma}^{\star}$ as a function of $\tau_{\rm{exp}}^3$ for different initial conditions:
$N_{\rm{i}}=5\cdot10^4$, $\bar{\rho}_{\rm{i}}=1.1\cdot10^9$cm$^{-3}$,
$T_{\rm{e}}=50$K;
$N_{\rm{i}}=4\cdot10^4$, $\bar{\rho}_{\rm{i}}=3\cdot10^9$cm$^{-3}$,
$T_{\rm{e}}=45$K;
$N_{\rm{i}}=5\cdot10^4$, $\bar{\rho}_{\rm{i}}=10^9$cm$^{-3}$,
$T_{\rm{e}}=33.3$K;
$N_{\rm{i}}=8\cdot10^4$, $\bar{\rho}_{\rm{i}}=10^9$cm$^{-3}$,
$T_{\rm{e}}=38$K;
$N_{\rm{i}}=10^5$, $\bar{\rho}_{\rm{i}}=1.3\cdot10^9$cm$^{-3}$,
$T_{\rm{e}}=33.3$K (left to right). The solid line is a linear fit. The error
bars show the range of two to eight percent relative deviation between
$\Gamma_{\rm i}$
and $\tilde{\Gamma}_{\rm i}$ for determining $t_{\Gamma}^{\star}$.}
\end{figure}

Equating $\tau_{\rm{corr}}$ and $\tau_{\rm{hyd}}$ as given above yields
\begin{equation}
\tau_{\rho}^{\star} = 2^{-1/2} \tau_{\rm{exp}} x^2 \sqrt{1+\sqrt{1+4x^{-4}}}
\approx \tau_{\rm{exp}} x^2
\end{equation}
with $x \equiv \tau_{\rm{exp}} / (3 \alpha)$ as the time
when both timescales become equal. In fig.\ \ref{fig4}, we show the time
$\tau_{\Gamma}^{\star}$ when correlations start to freeze out as a function of
$\tau_{\rm{exp}}^3$, where $\tau_{\Gamma}^{\star}$ is determined as the
time when the relative deviation between $\Gamma_{\rm i}$ and
$\tilde{\Gamma}_{\rm i}$ is less
than five percent for the last time. The linear correlation visible in the figure
strongly supports our reasoning that it is the cross-over of timescales
that is responsible for the freeze-out of correlations.

Thus, we may conclude that the system ultimately approaches a non-equilibrium
undercorrelated 
state again due to the correlation freeze-out described above. Still, the pair
correlation functions can well be fitted to those of an equilibrium plasma
at this stage (fig.\ \ref{fig3}(d)), in contrast to the behavior at early
times. This is due to the fact that the system went through a phase where
equilibrium spatial correlations have developed which are preserved during
the further evolution of the plasma. Hence, the system has the correlation
properties of an equilibrium system, however ``with the wrong temperature''.

In conclusion, we have simulated an expanding ultracold neutral plasma
with special attention to the formation of ionic correlations. We have found
that several phases can be distinguished in the evolution of the system.
First, a quick relaxation to local equilibrium occurs, together with its
characteristic transient oscillations of the ion temperature. After that,
the system is close to a --- changing --- local equilibrium. In this stage,
a CCP defined from temperature and density indeed is a measure
for correlations in the plasma. Moreover, and this has, to our
knowledge, not been pointed out so far, the plasma reaches a state
well inside the strongly coupled regime, with $\Gamma_{\rm i} \agt 10$.
Ultimately, the timescale for equilibration becomes longer than the timescale
on which the equilibrium changes, thus  the system cannot equilibrate anymore
and correlations freeze out. Clearly, ultracold neutral plasmas are unique
systems that evolve through different thermodynamical
stages of non-equilibrium and (near)-equilibrium behavior. Their further
experimental and theoretical study thus should provide new stimulus for
plasma physics as well as for non-equilibrium thermodynamics.

\end{document}